\documentclass[letterpaper]{article}
\usepackage{authblk}

\usepackage{geometry}
\usepackage{cite}
\usepackage{amsmath,amssymb,amsfonts}
\usepackage{algorithmic}
\usepackage{comment}
\usepackage{graphicx}
\usepackage{subcaption}
\usepackage{float}
\usepackage{todonotes}
\usepackage{textcomp}
\usepackage{xcolor}
\usepackage{multirow}
\usepackage[thinlines]{easytable}
\usepackage{makecell}
\usepackage[normalem]{ulem} 

\definecolor{darkgreen}{cmyk}{0.90,0,0.80,0.20}
\definecolor{darkred}{cmyk}{0, 1, 1, 0.10}
\definecolor{lightblue}{cmyk}{1, 0.70, 0, 0}

\usepackage{color}
\usepackage{soul}

\def\BibTeX{{\rm B\kern-.05em{\sc i\kern-.025em b}\kern-.08em
    T\kern-.1667em\lower.7ex\hbox{E}\kern-.125emX}}

\newcommand{\etal}{\textit{et al}.}

\begin{document}

\providecommand{\keywords}[1]
{
  \small	
  \textbf{\textit{Keywords---}} #1
}

\title{An Overview of Laser Injection against\\ Embedded Neural Network Models\\}

\author[1,2]{Mathieu Dumont}
\author[1,2]{Pierre-Alain Moëllic}
\author[3]{Raphael Viera}
\author[3]{Jean-Max Dutertre}
\author[1,2]{Rémi Bernhard}

\affil[1]{CEA Tech, Centre CMP, Equipe Commune CEA Tech - Mines Saint-Etienne, F-13541 Gardanne, France}
\affil[2]{Univ. Grenoble Alpes, CEA, Leti, F-38000 Grenoble, France\protect\\
\textit{\{mathieu.dumont,pierre-alain.moellic,remi.bernhard\}@cea.fr}}
\affil[3]{Mines Saint-Etienne, CEA Tech, Centre CMP, F - 13541 Gardanne, France\protect\\
\textit{\{raphael.viera,dutertre\}@emse.fr}}

\date{}

\maketitle

\begin{abstract}
For many IoT domains, Machine Learning and more particularly Deep Learning brings very efficient solutions to handle complex data and perform challenging and mostly critical tasks. However, the deployment of models in a large variety of devices faces several obstacles related to trust and security. The latest is particularly critical since the demonstrations of severe flaws impacting the integrity, confidentiality and accessibility of neural network models. 
However, the attack surface of such embedded systems cannot be reduced to abstract flaws but must encompass the physical threats related to the implementation of these models within hardware platforms (e.g., 32-bit microcontrollers). Among physical attacks, Fault Injection Analysis (FIA) are known to be very powerful with a large spectrum of attack vectors. Most importantly, highly focused FIA techniques such as laser beam injection enable very accurate evaluation of the vulnerabilities as well as the robustness of embedded systems. Here, we propose to discuss how laser injection with state-of-the-art equipment, combined with theoretical evidences from Adversarial Machine Learning, highlights worrying threats against the integrity of deep learning inference and claims that join efforts from the theoretical AI and Physical Security communities are a urgent need.  
\end{abstract}


\section{Introduction}
\label{introduction}
A major challenge in Machine Learning (ML) is the growing popularity of edge-deployed models and particularly deep neural networks in a large variety of embedded systems. High interest is shown on porting Deep Neural Networks (DNN) on constrained platforms, such as 32-bit low power microcontrollers, especially for inference purposes. However, this massive deployment of edge neural networks brings new security challenges that will be soon major issues for various critical domains, especially for those relying on IoT. Indeed, ML models are expected to be included in numerous devices, which can be hacked with an extensive overall attack surface, because of critical flaws intrinsically related to the ML algorithms as much as their implementation in physically accessible devices. These new threats could be critical for domains such as smart transport, autonomous driving, medical systems or cybersecurity infrastructures.

In the ML and Security state-of-the-art, adversarial examples \cite{szegedy2013intriguing} has emerged as a major threat against complex models such as deep neural networks. This phenomenon has disastrous impacts on the development, trust and durability of ML models by targeting their integrity~\cite{papernot2016}. Adversarial examples consist of input data which have been modified in a way that is intended to cause the ML to misclassify it. An interesting aspect of this attack is that the adversarial perturbation is minimal in magnitude or semantically or contextually coherent (e.g. a graffiti on a road sign), so that a human observer could hardly detect the perturbation.

On the other hand, works on physical attacks targeting embedded DNN are less common. However, these kinds of attacks constitute a real threat against embedded systems, as proven in other field, as smartcard security, in which it constitutes an essential topic for semiconductors companies and, therefore is included in certification schemes~\cite{lomne2016common}. A wide range of physical attacks~--~like Side-Channel Analysis (SCA) or FIA~--~and countermeasures has already been studied and developed mainly to secure cryptographic modules or authentication process~\cite{DPAbook, Barenghi2012}. Edge neural networks contains some critical data too, IP model, or personal data (banking information, medical records) and are designed to achieve critical tasks. Being included in a large variety of devices, the integrity and confidentiality of embedded DNN models can be threaten with attacks led directly on the Integrated Circuits.

First works on physical attack targeting embedded DNN models exploits SCA to reverse engineer a model~\cite{Batina2018,Xiang2020}.
As for SCA, we are at the prelude of fault injection techniques against embedded models. First experiments targeted several activation functions with laser injections as well as different attack paths~\cite{Hou2020}. Recently, each deep network layer are placed separately on a FPGA floorplan in order to characterize each one under a laser beam~\cite{Benevenuti2018}.
 
Those laser injection attacks are effective, but they generally involve large disruptions in DNN. Comparatively, adversarial examples require only small perturbation to fool a model. Therefore, analyzing the advantage and practicability of reproducing similar type of disturbance with a highly focus injection technique such as infrared laser beam is becoming an important concern.

In this work, we propose a panorama of fault injection techniques targeting embedded DNN with a particular attention on laser injection. We make a connection with theoretical evidences from adversarial machine learning to discuss the criticality of the different attack vectors present to an adversary (i.e. trying to answer the question \textit{where to shoot?}) whose objective is to strike the integrity of an embedded DNN. More particularly:  

\begin{itemize}
\item we highlight the \textit{criticality} of sparse adversarial examples that we affirm as being relevant in a fault injection context by targeting only few bits of the input (Section~\ref{adv_exp_fi});  
\item we demonstrate the power of laser fault injection on a typical IoT platform (32-bit microcontroller) with a cutting-edge equipment (Section~\ref{fia});
\item after reviewing the literature on fault injection against DNN models, we discuss the different strategies available for an adversary (Section~\ref{fia_ednn}). 
\end{itemize}

\section{Adversarial Examples}
\label{adv_exp_fi}
\subsection{Definition}
\label{definition_adv_exp}

A supervised neural network model $M_\theta$ is a parametric model that aims at mapping an input space $\mathcal{X}=\mathbb{R}^d$ (e.g., images, speech) to an output space $\mathcal{Y}$. 
For a classification task, $\mathcal{Y}$ is a finite set of labels $\{1,...,C\}$. $M$ is trained by minimizing a loss function $\mathcal{L}$ that quantifies the error between a prediction $\hat{y} = M_\theta(x)$ and the groundtruth label $y$. At inference time, when fed by an input $x \in \mathcal{X}$, the learned model outputs a set of raw predictions (the \textit{logits}) that are mapped in a probabilistic formalism thanks to a softmax function in order to provide the probability $p(l_i|x)$ for each label $l_i \in \mathcal{Y}$.   

Neural network models have been shown to be vulnerable to several integrity, confidentiality and accessibility-based threats. Adversarial examples \cite{szegedy2013intriguing} are probably the most studied integrity-based attacks with numerous attacks demonstrated in both white and black-box settings. Adversarial examples are defined as maliciously modified inputs that fool a model at inference time. Formally, a classical definition of an adversarial example $x'$, crafted from a clean input $x$, defines the adversarial perturbation with the $l_{p}$ norm: 
\begin{align} 
\label{eq_adv_def}
\begin{gathered}
\min_{x'} ||x-x'||_{p} 
  \text{  s.t. } x' \in \mathcal{X},
  M_{\theta}(x') = l', 
  l' \neq M(x)
\end{gathered}
\end{align}

The attack can be \textit{targeted} ($l'$ is chosen by the adversary) or \textit{non-targeted}.
Many directions have been explored to explain and characterize this phenomenon \cite{ilyas2019adversarial} and defenses are regularly proposed to improve the robustness of models against adversarial perturbation, with a particular focus on white-box setting (i.e. considering an adversary with a full access and knowledge of the target model)~\cite{Madry2017}. Black-box attacks have been efficiently proposed even in very restrictive settings with attackers having access only to the predicted label~\cite{cheng2019}.   

\subsection{Sparse adversarial perturbations}
\label{l0}

Equation~\ref{eq_adv_def} defines the classical $l_{p}$ norm setting for adversarial examples. Usually, the adversary aims at crafting \textit{imperceptible} perturbation which is bounded by an \textit{adversarial budget}, $||x-x'||_{p}\leq\epsilon$. Classically, three norms are used: $l_2$ is the classical euclidean norm,  $l_\infty$ constrains  the maximum change to any dimension of the input (e.g. pixel) and a $l_0$-based attack aims at minimizing the number of perturbed dimensions. Thus, for classical $l_2$ or $l_\infty$ attacks, every dimension is altered whereas $l_0$ attacks only target very few dimension of $x$. Interestingly, some state-of-the-art attacks, such as Carlini \& Wagner~\cite{carlini2017towards} and PGD~\cite{Madry2017} encompass a generic $l_p$ framework. 

In the context of fault injection analysis, $l_0$ attacks are particularly interesting since they match a practical objective by altering a minimal set of dimensions. The Jacobian-base Saliency Map Attack (JSMA) method~\cite{papernot2016} was one of the first approaches dedicated to sparse perturbations by only altering the dimensions that have high saliency (Jacobian matrix of the model). Then, Su \etal~propose an \textit{optimal} $l_0$ attack for image classification by altering a single pixel with the so-called ''One-Pixel Attack'' (OPA)~\cite{su2019one}. The authors use an evolutionary algorithm (differential evolution~--~DE) to find the best pixel ($x,y$) and perturbation ($r$, $g$, $b$ for a color image). Although the approach can be generalized to more than \textit{one} pixel, they mainly focus their experiments on this objective. 
Because of the strict constraints of the threat model, the success rate of the attack is around 30\% on classical image benchmarks (CIFAR10) contrary to state-of-the-art $l_\infty$ or $l_2$ gradient-based attacks that reach 100\%.     

\begin{figure}[t]
    \centering
    \includegraphics[width=\linewidth]{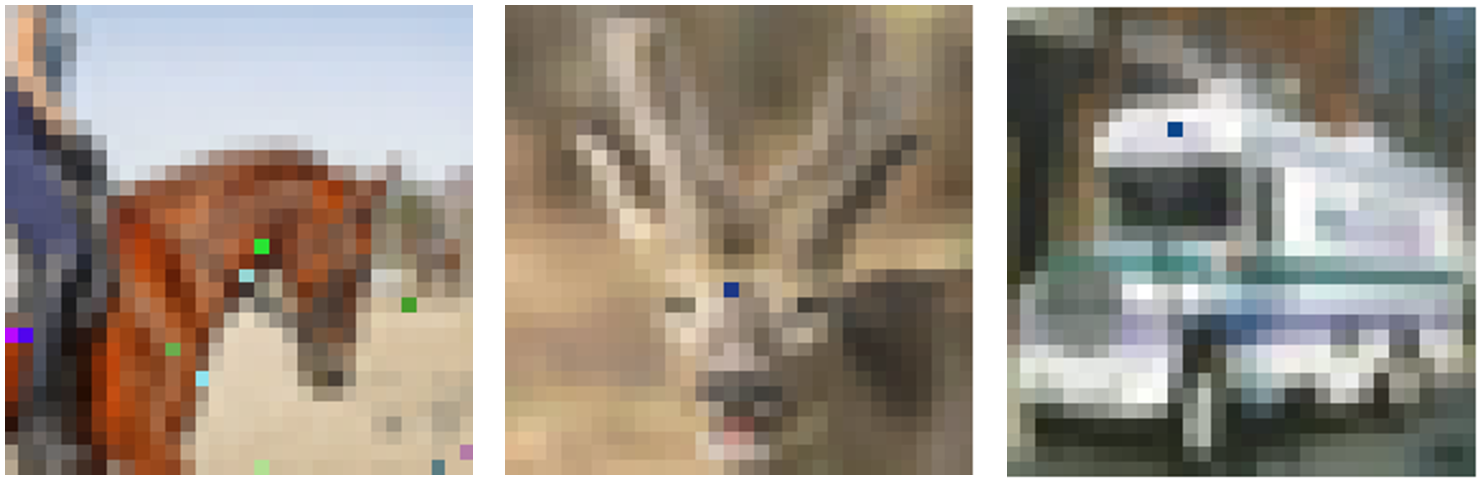}
    
    \caption{CIFAR10 images correctly predicted (\textit{horse}, \textit{deer}, \textit{truck}) by a ResNet model. Left: OPA selects a set of 10 sensitive pixels. These pixels are exhaustively tested with bit-flip perturbation. Middle and Right: a successful one bit-flip attack on a single pixel that respectively fools the model to \textit{cat} and \textit{ship}.}
    \label{one_bit_attack}
\end{figure}

Other approaches propose to improve the success of sparse perturbation by considering more than one dimension to be changed as well as combining sparsity and imperceptibility (e.g. by mixing $l_0$ and $l_\infty$ constraints) such as~\cite{croce2019sparse} that reach more than 95\% success rate on CIFAR10 by perturbing at most 10 pixels. Moreover, because the $l_0$ constraint generally leads to complex optimization problems, SparseFool~\cite{modas2019sparsefool} proposes to add sparsity in a state-of-the-art iterative $l_2$ attack (DeepFool) by linearly approximating the decision boundary.

By using the basic principle of typical $l_0$ attacks, we can demonstrate that even a single bit-level perturbation can lead to a successful adversarial attack. We illustrate that point on the classical CIFAR10 data set, with a state-of-the-art ResNet model. The OPA from~\cite{su2019one} or CornerSearch from~\cite{croce2019sparse} enable to select a set of sensitive pixels. According to a fault model (bit-flip or bit-set model), these pixels are exhaustively faulted and the resulting images are tested through the model. Figure~\ref{one_bit_attack} (left) shows a \textit{horse} image and 10 sensitive pixels selected by the OPA. An exhaustive search shows that 19 bit-flip from these pixels effectively lead to a misclassification. The One-Bit-Attack is illustrated with the middle and right images, the model is effectively fooled with a single bit-flip. A detailed evaluation of such a "One-Bit-Attack" is out of the scope of this review paper and will be proposed in further publications. Here, the important outcome is that theoretical adversarial machine learning clearly expose attack vectors that may be efficiently exploited in a physical attack context and, therefore, increase the dangers of an adversary having a physical access to an embedded DNN model.

\section{Fault Injection Analysis}
\label{fia}
As mentioned in the introduction, the attack surface for the integrity of an embedded DNN model cannot be limited to algorithmic threats such as the aforementioned adversarial examples. The threat model must encompass physical flaws related to the implementation of the model in a particular device. In this section, we detail and experimentally demonstrate how state-of-the-art fault injection can severely impact a typical 32-bit microcontroller platform used in many IoT domains.   

\subsection{Fault Injection Analysis}
Fault injection analysis consists in disrupting the circuit computation in order to induce a faulty behavior whose observation can be exploited to extract confidential information. In cryptography, fault injection-based attacks represent a major way to recover cryptographic key from an AES~\cite{Barenghi2012}. A classical approach is to perform a Differential Fault Analysis (DFA) which compare two sets of encryption results (correct and faulted ciphers) that can be efficiently linked to some hypothesis on the secret key. Moreover, fault injections are known to be very effective to break authentication routines such as PIN verification on smartcard or bypass the authentication process of a secure bootloader. 

A wide range of FIA have been demonstrated~\cite{Barenghi2012} depending on the adversary's goal, expertise, budget as well as the target characteristics. \textit{Low cost} techniques can simply rely on glitching the power supply or clock of the target while other methods rely on more focus and advanced techniques such as ElectroMagnetic (EM) Fault Injection, Forward Body Biased Injection or Laser Fault Injection. Although the latter can be more expensive depending on the equipment, it offers a superior temporal and spatial resolution, which makes it possible to fault a single bit during circuit operation. 
Because a sound security evaluation must encompass \textit{worst-case scenario} with advanced threats, laser-based injection analysis is usually considered as an important evaluation framework.  

\subsection{Laser Fault Model}
\label{laser_sota}

Local fault injection techniques are used to disturb specific parts of a microarchitecture without affecting others. 
A laser injection enables an attacker to induce single bit-flips in static memory cells. The SRAM cell is a well-known target for this kind of attack and \textit{bit-set/reset} fault \cite{Roscian2013} can be obtained on this IC part.   
Recent studies \cite{Menu2020Sing} \cite{Colombier2019}, show that the array of the NOR Flash memory of a 32-bits microcontroller is sensitive to laser fault injection. Transient \textit{bit-set} faults are induced during the read operation of the Flash memory. Moving along one axis of the flash memory area allows to precisely target the bits of the fetched data one after the other. One of the advantages is that the data stored in Flash memories remain unaltered. 
The authors succeed to bypass a PIN verification and to recover an AES key. This fault model is reproduced on our laser bench in the next section.

\subsection{Experimental Setup}
To demonstrate the high precision of laser fault injection, we perform some experiments on a typical platform for many IoT domains, namely a 32-bit microcontroller.  

\begin{figure}[t]
    \centering
    \includegraphics[scale=0.70]{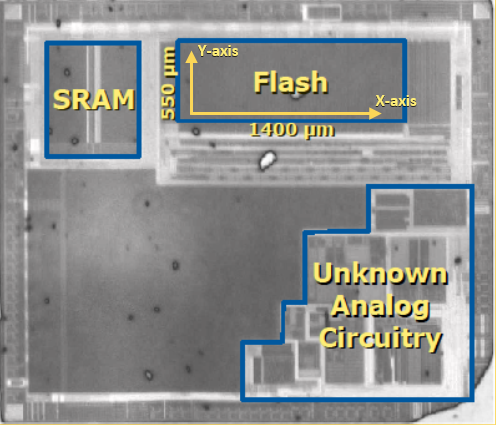}   
    \caption{Infrared picture of the target 32-bit microcontroller.} 
    \label{STM32}
\end{figure}

\textbf{Target board and microcontroller.}
The targeted chip in our experiment embeds an ARM Cortex-M3 running at 7.4\,MHz. It includes 128kB of Flash memory and is manufactured in the 90\,nm technology node. An infrared picture of the target microcontroller is shown in Figure \ref{STM32}. The chip dimension is 3mm x 2.5mm. Main parts of the chip are outlined including the Flash memory which constitutes the target of our attacks. The chip is opened on the back-side to provide an access to its laser-sensitive parts.

\textbf{Laser platform.} 
Our laser fault injection platform integrates two independent near infrared laser spots (focused through the same lens) with a wavelength of 1,064\,nm. Each laser spot has a diameter ranging from 1.5 µm to several tens of micrometers. The two spots can move independently inside the whole field of view of the lens with minimum distortion. The laser source provides a maximum power of 1,700 mW (measured by a photo-diode). 
The delay between the trigger and the laser shot can be adjusted with a step of a few ns. An infrared camera is used to focus the laser spot with regard to the target and a $XY$ stage enables to move the objective. 

\textbf{Assembly test code.}
To characterize the Single-bit fault model we used a simple flash reading code. 32-bit word are used to highlight \textit{bit-set} (0x00000000) and \textit{bit-reset} (0xFFFFFFFF). The executed code fetches the 32-bit word with a LDR (load) instruction and stores it in the R0 register. The LDR instruction is surrounded by 16xNOP before and after to prevent instruction corruption. The laser pulse is synchronized shortly after the decoding of the LDR to target the clock cycle where the data is actually read. 

\subsection{Single-Bit fault on Flash Memory}
\label{Single-bit}
\begin{figure}[t]
    \centering
    \includegraphics[width=0.8\linewidth]{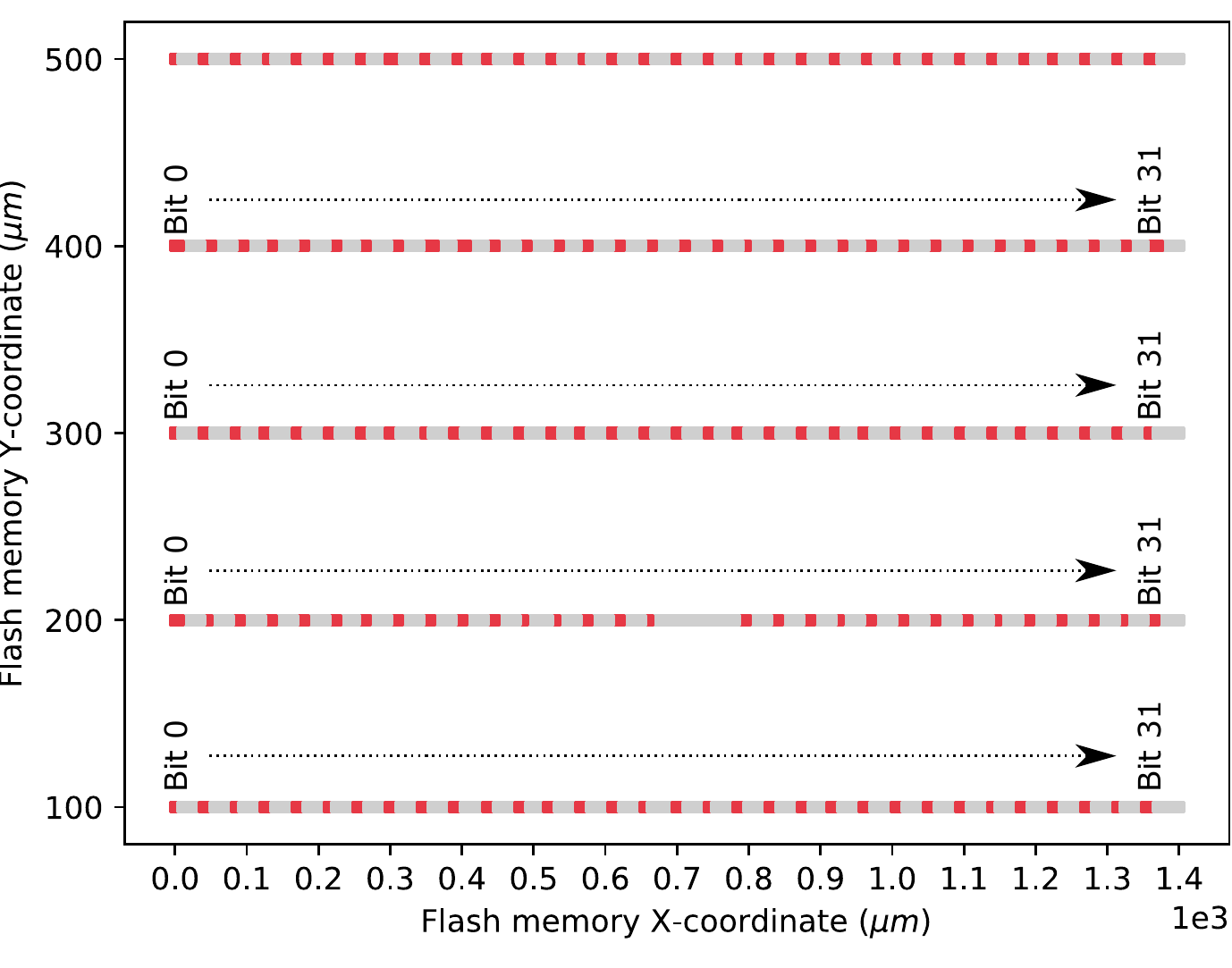}
    \caption{Mapping scan of the microcontroller embedded Flash memory. Induced bit-set from bit 0 to bit 31 depending on the X position.}
    \label{One_spot}
\end{figure}

For this first experiment, to retrieve the single \textit{bit-set} fault model, we only use one laser spot. The x5 optical lens is chosen which corresponds to a laser spot size of 15\,µm. A mapping scan is performed on the Flash area with a dimensions of X=1400\,µm and Y=550\,µm. The elementary step along the X-axis is 5\,µm and 100\,µm for the Y-Axis. The optimal parameters for inducing faults were measured with a minimum pulse power of 200\,mW, a pulse width of 200\,ns and a delay of 1,700\,ns.
Figure \ref{One_spot} gives the mapping scan of the Flash memory under laser beam. We observed that moving along the X-axis, every bit of the 32-bit register is faulted one after the another as well as in \cite{Menu2020Sing}. On the contrary, moving along Y-axis has no effect on the faulted-bit. Then, depending on the X position, a single \textit{bit-set} fault could be induced on a targeted bit. We were not able to performed \textit{bit-reset}. Some points are missing at X=750\,µm and Y=200\,µm due to some residual packaging plastic on the Flash area which obstruct the laser beam. This experiment shows that predictable and repeatable \textit{bit-set} fault can be induced in Flash with a bit-accurate precision.   

\subsection{Experiments with two laser spots}
\label{double_spot}

Double-spot laser platforms are not very common. Nevertheless, it provides the possibility to hit two different locations on the chip, at the same time.

We performed the same experiment than the previous one, however the Y-position is fixed to 100\,µm to focus on the X-Axis motion and the Spot 1 Laser position is set at X=0\,µm. At each iteration, the Spot 1 induced a bit-set at the fixed position of 0 µm while the Spot 2 will move along the X-Axis.

\begin{figure}[t]
    \centering
    \includegraphics[width=0.8\linewidth]{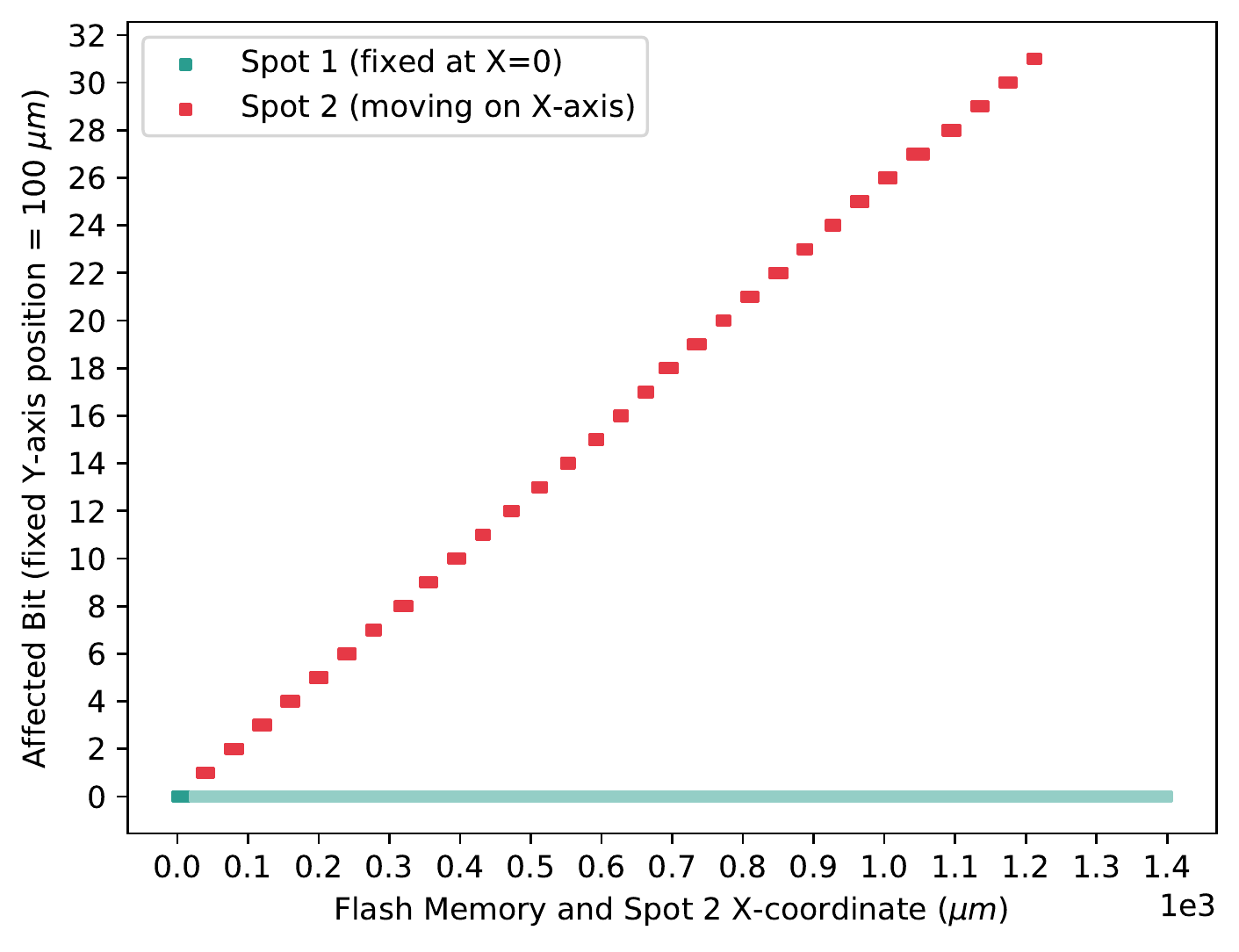}
    \caption{Mapping scan of the two laser spots on the microcontroller Flash memory. Spot 1 is fixed at X=0 µm, while Spot 2 is moved along the X-Axis. At each iteration, two faults are induced simultaneously, Spot 1 induced bit-set on bit 0 while Spot 2 induced bit-set from bit 1 to 31 depending on the X position.}
    \label{Double_spot}
\end{figure}

Results from the experiment are shown in Figure \ref{Double_spot}. The X-Axis correspond to the Spot 2 motion on the Flash area while the Y-Axis provided the associated faulted bit from bit 0 to bit 31. We observe that two \textit{bit-set} faults are induced simultaneously on two different bit value of the register. For example with Laser Spot 1 and 2 at the respective position of 0 µm and 700 µm, the data read by Flash memory is 0x00020001, meaning that bit n°1 and 18 have been set to logical value '1'.

With these experiments, we show that Laser Fault Injection could induce \textit{bit-set} in Flash memory with high temporal and spatial resolution. By demonstrating the accuracy and repeatability of this attack, we open new possibilities of stealthiness physical attack against neural network (since DNN parameters are in the Flash memory) in a close way of \textit{Sparse Adversarial} perturbations. Although further communications will address this new attack, we discuss the impact of such attack vector in the Section~\ref{perspective}.

\section{FIA for Embedded Neural Networks}
\label{fia_ednn}

\begin{table*}[t]
 \centering
 \resizebox{\textwidth}{!}{%
 \begin{tabular}{cccccc}
  \hline
  \textbf{\makecell{Fault Injection\\ Attack}} & \textbf{\makecell{Parameter\\Targeted}} & \textbf{\makecell{Layer\\Targeted}} & \textbf{\makecell{Simulation/\\Experimental}} & \textbf{\makecell{Order of magnitude of the number\\of faulted parameters/bit}}\\
  
  \hline
  \hline
  SBA \cite{Liu2017Fault} & Bias & \multirow{2}{*}{Hidden and Output} & Simulation & \#1 \textbf{(1)} \\
  GDA \cite{Liu2017Fault} & Weights &  & CNN (MNIST, CIFAR10) & \#1000 \textbf{(2)} \\
  
  \hline
  \multirow{2}{*}{BFA \cite{rakin2019bit}} & \multirow{2}{*}{Weights} & \multirow{2}{*}{All} & Simulation & \multirow{2}{*}{\#10 Bit-flip \textbf{(3)}} \\
   &  &  & ResNet models (CIFAR10, ImageNet) &  \\
  
  \hline
  &  &  & Both  & Single Fault Strategy: \#1 if exist \\
  DeepLaser* \cite{Hou2020} & Activation functions & Last hidden layer & MLP with 4 hidden layer (MNIST) & Multiple Fault Strategy: up to 40\% of faulted \\
  &  &  & Laser on activation function primitive & neurons for 50\% misclassification rate\\

  \hline
  
 \end{tabular}}
 \caption{Summary of fault injection attacks against DNN models mentioned in the paper. MLP is for Multi-Layer Perceptron. \textbf{(1)} The accuracy impact of SBA depends on the layer, see~\cite{Liu2017Fault}. \textbf{(2)} The number of impacted weights strongly depends on the layer, see~\cite{Liu2017Fault}. \textbf{(3)} For BFA, the attack success corresponds to a top-1 accuracy below random guess. (*) We use ''DeepLaser'' as it is the name of the initial work from Breier \etal}
  \label{attack_summarize}
\end{table*}

\subsection{Review}

Faults injection applied to neural network is a recent topic. We highlight some important works that pave the way to further studies and experiments on this field and summarize them in the Table~\ref{attack_summarize}.   
As a first milestone, Liu \etal~\cite{Liu2017Fault} theoretically introduces two types of attacks, the \textit{Single Bias Attack} (SBA) and the \textit{Gradient Descent Attack} (GDA) that are not practically evaluated with a fault injection method (even if laser fault injection and the memory-based Row-Hammer attack are mentioned as potential means). SBA aims at altering the bias of an output neuron or a neuron in a hidden layer. In the last case, the attack is only possible with a network that uses the ReLu activation function and require a severe alteration of the bias value. Interestingly, GDA uses gradient information to evaluate the influence of a perturbation (i.e. a fault) applied on a parameter on the output probability of a specific label. Since, this gradient-based approach leads to too many impacted parameters, a compression method is proposed to reduce the number of faults to apply. However, for a typical ResNet architecture on the CIFAR10 data set, GDA still requires hundreds of faults to be efficient. 

However, Rakin \etal~\cite{rakin2019bit,Rakin2020TBFA} highlight a major limitation to this work since it considers full precision floating-point models which significantly facilitates the attack and does not fit with current practices in embedded neural network models that support 8-bit quantized (or even binary) models. Actually, they show that a single bit-flip of the most significant exponent bit of a
random full-precision weight completely threaten the integrity of a ResNet model trained on ImageNet. Thus, they propose a novel DNN weight attack methodology called \textit{Bit-Flip Attack} (BFA) that combined the crafting principle of adversarial example and a progressive search procedure to identify the most vulnerable bits of the 8-bit weights. 
They succeed to degrade the accuracy of a ResNet-18 DNN from 69.8\% to 0.1\% with only 13 bits-flips out of 93 million bits. However, the practicability of the attack with a real fault injection means is not proposed. 

Safety-based simulation frameworks could also be used to assess the resilience of DNN by analyzing the sensitivity of the different parts of a model to random faults. ARES~\cite{Reagen2018ARES} highlights the vulnerability of DNN, by inducing static or transient hardware faults during the DNN inference. TensorFI~\cite{Chen2020Tensor} induces both hardware and software faults in general TensorFlow programs by duplicating the TensorFlow graph and creating a parallel Fault Injection graphs. Both show that even though DNN models trained with state-of-the-art procedure (e.g., regularization, data augmentation) have an intrinsic robustness against noise and random alterations, this robustness differs according to their architecture (convolutional layers, depth...) and that activation layers are less sensitive than weights. 

Although mentioning laser injection, clock glitching or Row-Hammer attacks, all of the previous works have been evaluated through simulations and (to the best of our knowledge) their practicability with real fault injection means has not been demonstrated yet. Therefore, some works endeavour to practically experiment and demonstrate fault injection on model embedded in hardware platforms such as microcontrollers.   
In \cite{Hou2020} faults are injected with a laser into several activation function primitives implemented on a ATmega328P microcontroller (8-bit). Experiments show that, depending on the targeted activation function, a neuron can be rendered inactive with a skip fault instruction, or the function output value could be altered. Then simulations evaluate the impact of such faults on the last hidden layer of a 4-layers network trained on the MNIST data set, when considering two strategies: the \textit{single fault} strategy has no guarantee of existence, the \textit{multiple fault} strategy requires up to 40\% of faulted neurons to reach a 50\% misclassification rate. 

\subsection{Perspective and Discussions}
\label{perspective}
The previous review shows that an adversary has several options at his disposal to alter the model's prediction with fault injection. However, an important aspect of the threat model is related to the adversary's capability and the way he can interact with the system. Indeed, in many IoT-related domains, inputs are directly captured by sensors (e.g. a camera) and, therefore, are very distinct to typical workflow used in adversarial examples works, such as ML-as-a-service API, where users \textit{directly} provide input data. 

A first logical strategy is to focus the faults on the final layers, i.e. by targeting the logits (penultimate layer) or the final softmax outputs. As observed in~\cite{Hou2020}, the softmax function is not easily altered contrary to other activation functions but, because softmax is only a normalization mapping and does not impact the order of the predictions, an adversary could have a better alternative by targeting the weights or the bias of the last hidden layer. However, from a practical point of view (unfortunately hardly taken into consideration in the literature), the inference forward pass is far from being a time-constant process, therefore an efficient injection \textit{at the end} of the network is very challenging.

A second option relies on the internal hidden layers. However, as noticed in safety-based works~\cite{Reagen2018ARES}, because of the intrinsic robustness of models (more particularly with data augmentation or regularization-based techniques such as dropout) an important amount of faults have to be achieved in order to mislead the model.    

The last alternative is to target the input layer by using the theoretical evidences highlighted by adversarial examples. Considering our previous remark above on typical sensor-based systems, we claim that this option is of high interest for an adversary. In a well-trained DNN, a random change in an input data should have no impact on the prediction. Then, the challenge is to fool the model with only a few faulted-bits of the input. As demonstrated in Section~\ref{l0}, it is what we can learn from state-of-the-art works that have demonstrated the efficiency of sparse adversarial examples. 

The experiments in Section~\ref{Single-bit} show that striking the internal parameters of a model as well as the opcode of the instructions of the inference process can be achieved accurately. However, SRAM is also known as being vulnerable against fault injection~\cite{lacruche2015}. For a typical 32-bit microcontroller such as ones with a Cortex M4 or M7 core, the model parameters are stored in the Flash memory while the SRAM handles the intermediate results (i.e. activation outputs). Therefore, we highlight the fact that major vulnerabilities related to accurate fault injection means concern the inference flow in its entirety, even with single fault at the bit-level (crafted through lessons learned from theoretical adversarial machine learning) that could be hardly detectable.

\section{Conclusion}
\label{conclusion}
This paper deals with a major concern against the integrity of embedded DNN models that are increasingly deployed in many IoT devices. Because an adversary has a physical access to these devices, the attack surface encompasses theoretical flaws (such as adversarial examples) \textit{and} physical implementations threats (such as FIA).
We show that FIA-based threats are now a reality for embedded machine learning models, more particularly, DNN. Among the fault injection techniques, laser injection is a powerful mean to assess the robustness of models with very focused faults as we demonstrated the impact on 32-bit microcontroller, a typical IoT platform. A panorama of the proposed attacks show that this topic is a growing concern for the Security and the Artificial Intelligence community, even if the majority of the works are limited to simulation efforts or do not take into consideration the characteristics of embedded models (e.g. low bitwidth parameters). By reviewing state-of-the-art fault injection attack vectors against DNN model, we claim that lessons could be learned from theoretical adversarial machine learning to better foresee future advanced threats and design proper defense schemes.

\section*{Acknowledgments}
This work is a collaborative research action that is partially supported by (for \textbf{CEA-Leti}) the European project ECSEL InSecTT\footnote{\url{www.insectt.eu}, InSecTT: ECSEL Joint Undertaking (JU) under grant agreement No 876038. The JU receives support from the European Union’s Horizon 2020 research and innovation program and Austria, Sweden, Spain, Italy, France, Portugal, Ireland, Finland, Slovenia, Poland, Netherlands, Turkey. The document reflects only the author’s view and the Commission is not responsible for any use that may be made of the information it contains.} and by the French National Research Agency (ANR) in the framework of the \textit{Investissements d’avenir} program (ANR-10-AIRT-05, irtnanoelec);  and supported (for \textbf{Mines Saint-Etienne}) by the French funded ANR program PICTURE (AAPG2020).
Both partners thank the MicroPacks platform for their support and cutting-edge equipments. 

\bibliography{bib/biblio_1, bib/biblio_2}
\bibliographystyle{ieeetr}

\end{document}